\documentclass[aps,twocolumn,nofootinbib]{revtex4}
\usepackage[latin1]{inputenc}
\usepackage{epsfig}
\newcommand{\beq}{\begin{equation}}
\newcommand{\eeq}{\end{equation}}

\begin{document}

\title{Entropy production of the contact model}

\author{Tânia Tomé and Mário J. de Oliveira}
\affiliation{Universidade de São Paulo, Instituto de Física,
Rua do Matão, 1371, 05508-090 São Paulo, SP, Brazil}

\begin{abstract}

We propose an expression for the production of entropy 
for system described by a stochastic dynamics which is
appropriate for the case where the reverse transition rate
vanishes but the forward transition is nonzero. 
The expression is positive definite and based on the
inequality $x\ln x-(x-1)\ge0$. The corresponding entropy
flux is linear in the probability distribution allowing
its calculation as an average. The expression is applied
to the one-dimensional contact process at the stationary
state. We found that the rate of entropy production per
site is finite with a singularity at the critical point
with diverging slope.

\end{abstract}

\maketitle

\section{Introduction}

The entropy production is the central concept of the theory of
systems out of thermodynamic equilibirum. This quantity characterizes
the state of nonequilibrium and is nonzero not only in systems that
have not yet reached thermodynamic equilibrium but also in systems
that are in a permanent nonequilibrium stationary state.
Within the stochastic approach to thermodynamics the production
of entropy is related to the ratio between the probability of a
trajectory and its reverse. 

Let us consider a stochastic dynamics in continuous time
and discrete space of states and let
$P_i$ be the probability distribution of states $i$, and 
$W_{ij}$ be the transition probability rate from state $j$
to state $i$. The equation governing the time evolution of $P_i$ is
assumed to be given by a master equation \cite{kampen1981,tome2015L}
\beq
\frac{dP_i}{dt} = \sum_j (W_{ij} P_j - W_{ji} P_i).
\label{3}
\eeq
The total production of entropy is a sum of terms 
each one corresponding to a transition $j\to i$, and given
by the Schnakenberg expression \cite{schnakenberg1976}
\beq
\Pi_{ij} =
\frac12(W_{ij}P_j - W_{ji}P_i) \ln\frac{W_{ij}P_j}{W_{ji}P_i}.
\label{6}
\eeq
An identical expression is valid for the reverse transition
$i\to j$. The expression (\ref{6}) is clearly nonnegative
because if we let $W_{ij}P_j/W_{ji}P_i=x$ then this expression
is proportional to
\beq
(x-1)\ln x \ge 0,
\eeq
which is always nonnegative,
reflecting the main feature of the production of entropy.
This is the usual expression for the rate of entropy production
and has been extensively studied and applied
\cite{luo1984,mou1986,lebowitz1999,maes2003,seifert2005,%
tome2006,zia2007,andrieux2007,harris2007,gaveau2009,esposito2009,%
szabo2010,tome2010,hinrichsen2011,esposito2012,tome2015,tome2018}.
In particular it has been used to determined the rate of
entropy production of irreversible stochastic lattice models 
\cite{crochik2005,tome2012,noa2019,tome2023,hawthorne2023}
at the nonequilibrium
stationary state.

The expression (\ref{6}) however breaks down when 
one transition rate vanishes while the other remains
nonzero, that is, when a transition has no reverse.
There are many examples of stochastic systems where this
happens, the most striking examples being the systems with
absorbing states such as the contact process
\cite{marro1999,hinrichsen2000,henkel2008}.
In this model, particles residing on a lattice are annihilated
and created according to the following rules. A particle is
annihilated spontaneously, and is created
in a empty site if this site has at least one neighboring site
occupied. Thus the transition corresponding to the 
annihilation of an isolated particle has no reverse,
and the application of the formula (\ref{6}) would result
in a divergent rate of entropy production.

It should be pointed out that the Schnakenberg formula
has a straight relationship with the Jarzynski equality.
Thus the breakdown of this formula results in the 
inapplicability of the Jarzynski inequality when 
the probability of the backward probability vanishes
wereas the forward probability remains finite \cite{murashita2014}.

This problem has been addressed by several authors
and there have been some attempts to overcome it
\cite{murashita2014,ohkubo2009,benavraham2011,zeraati2012,rahav2014,saha2016,
pal2017,busiello2020,pal2021,pal2021a}. Some of
these attempts focused mainly on the introduction of an
effective backward transitions while keeping 
formula (\ref{6}) unchanged. A distinguishing formula
was proposed by Busiello, Gupta, and Maritan
\cite{busiello2020} and will be commented further on.
Here we take a different route.
To determine the rate of entropy production when
the reverse transition is absent,
that is, when $W_{ij}> 0$ but $W_{ji}=0$,
we propose the following expression
\beq
\Pi_{ij} =
W_{ij}P_j\ln\frac{P_j}{P_i} - W_{ij}(P_j-P_i),
\label{7}
\eeq
which is associated to the transition $j\to i$. The production
associated to the reverse transition $i\to j$ is defined as
being zero. We see
that the expression (\ref{7}) is clearly nonnegative because
if we let $P_j/P_i=x$ then this expression is
proportional to
\beq
x\ln x - (x-1) \ge0.
\eeq

The present approach is applied to the calculation
of the rate of entropy production of the one-dimensional contact
process which is known to undergo a phase transition from 
an active to an absorbing state as one increases the annihilation
rate constant. The rate of entropy production
is found to be finite for any value of the annihilation parameter
displaying a singularity at the critical point.

\section{Entropy production}

The entropy $S$ of a system described by the master equation
(\ref{3}) is given by the Gibbs expression
\beq
S = - \sum_i P_i\ln P_i.
\eeq
Deriving $S$ with respect to time we get 
\beq
\frac{dS}{dt} = - \sum_i \frac{dP_i}{dt}\ln P_i.
\eeq
Using the master equation (\ref{3}) we reach the
expression 
\beq
\frac{dS}{dt} = \sum_{ij} W_{ij} P_j \ln \frac{P_j}{P_i},
\label{14}
\eeq
which is well defined no matter if $W_{ij}$ vanishes or not.

The total rate of entropy production $\Pi$ is understood as a
sum of terms $\Pi_{ij}$, 
\beq
\Pi = \sum_{ij}\Pi_{ij},
\eeq
each one associated to a transition $j\to i$ and defined as follows.
If $W_{ij}$ and its reverse $W_{ji}$ are nonzero, 
$\Pi_{ij}$ is given by the expression (\ref{6}).
If $W_{ij}$ is nonzero but the reverse transition rate
$W_{ij}$ vanishes, it is given by the expression (\ref{7}).

If we add $dS/dt$ and $\Pi$, we find  the total entropy flux
$\Psi$, that is, $\Psi = dS/dt + \Pi$, or
\beq
\frac{dS}{dt} = \Pi - \Psi,
\eeq
that is, the time variation of the entropy of a system is equal
to the rate of entropy production minus the flux of entropy
{\it from} the system {\it to} the outside.

Writing $\Psi$ as a sum of terms
\beq
\Psi = \sum_{ij} \Psi_{ij},
\eeq
we get the expressions for $\Psi_{ij}$,
understood as the flux of entropy associated to a
transition $j\to i$. It is given by
\beq
\Psi_{ij} = \frac12(W_{ij}P_j - W_{ji} P_i)
\ln\frac{W_{ij}}{W_{ji}},
\label{11}
\eeq
if $W_{ij}$ and its reverse $W_{ji}$ are nonzero, and by
\beq
\Psi_{ij} = - W_{ij} (P_j - P_i),
\label{12}
\eeq
if $W_{ij}$ is nonzero but the reverse transition
$W_{ij}$ vanishes. In this case we set $\Psi_{ji}=0$.
Both expressions (\ref{11}) and (\ref{12}) are linear
in $P_i$ so that the entropy flux can be understood
as an average over the distribution of probability
$P_i$.

In the stationary state $dS/dt$ vanishes and
$\Pi=\Psi$. If the stationary state is a nonequilibrium state
then $\Pi=\Psi>0$ and entropy are constantly being
produced and thrown away into the environment as
$\Psi>0$.

\section{Stochastic lattice models}

Stochastic lattice models are defined on a lattice
where each site $i$ holds a stochastic variable
$\sigma_i$ that takes discrete values. We consider
the simplest case where it takes only two values,
0 and 1, interpreted as the site being
empty or occupied by a particle. We denote by
$\overline{\sigma}_i$ the variable related to $\sigma_i$
in such a way that when the latter takes one of the
two values, the former takes the other.
The whole state of the system
is given by all $\sigma_i$ and is denoted by $\sigma$,
that is, 
\beq
\sigma = (\sigma_1, \sigma_2, \ldots, \sigma_i, \ldots).
\eeq
It is convenient to define the state $\sigma^i$, by
\beq
\sigma^i = (\sigma_1, \sigma_2, \ldots, \overline{\sigma}_i, \ldots).
\eeq

We restrict ourselves to models that changes only the state 
of one site. The transition probability of 
changing from $\sigma$ to $\sigma^i$ is denoted by
$w_i(\sigma)$ and the master equation that governs the
time evolution of the probability distribution $P(\sigma)$
is given by
\beq
\frac{d}{dt}P(\sigma) = \sum_i \{w_i(\sigma^i)P(\sigma^i)-w_i(\sigma)P(\sigma)\}.
\eeq

The entropy is given by
\beq
S = -\sum_\sigma P(\sigma)\ln P(\sigma),
\eeq
and its time variation is
\beq
\frac{dS}{dt} = \sum_i \sum_\sigma 
w_i(\sigma)P(\sigma)\ln \frac{P(\sigma)}{P(\sigma^i)},
\eeq
The rate of entropy production $\Pi$ is a sum of terms,
\beq
\Pi = \sum_i \Pi_i,
\eeq
each one given by
\beq
\Pi_i =  
\frac12\sum_\sigma\{w_i(\sigma)P(\sigma) - w_i(\sigma^i)P(\sigma^i)\}
\ln \frac{w_i(\sigma)P(\sigma)}{w_i(\sigma^i)P(\sigma^i)},
\eeq
if $w_i(\sigma)$ and $w_i(\sigma^i)$ are both nonzero, and
\beq
\Pi_i = \sum_\sigma 
w_i(\sigma)\{P(\sigma)\ln \frac{P(\sigma)}{P(\sigma^i)}
- P(\sigma) + P(\sigma^i)\},
\eeq 
if $w_i(\sigma)$ is nonzero but $w_i(\sigma^i)$ vanishes.

In an analogous manner, the entropy flux $\Psi$ is a sum of terms,
\beq
\Psi = \sum_i \Psi_i,
\label{22}
\eeq
each one given by
\beq
\Psi_i =  
\frac12\sum_\sigma \{w_i(\sigma)P(\sigma) - w_i(\sigma^i)P(\sigma^i)\}
\ln \frac{w_i(\sigma)}{w_i(\sigma^i)},
\label{9}
\eeq
if $w_i(\sigma)$ and $w_i(\sigma^i)$ are both nonzero, and
\beq
\Psi_i = - \sum_\sigma w_i(\sigma) \{P(\sigma) - P(\sigma^i)\},
\label{10}
\eeq
if $w_i(\sigma)$ is nonzero but $w_i(\sigma^i)$ vanishes.

\section{Contact model}

We apply the results above to a specific stochastic lattice
model, namely, the one-dimensional contact
process. The model is defined on a chain where each site
can be occupied by a particle or empty. 
The probability of
creating a particle in an empty site is equal to the fraction of
occupied neighboring sites, and the probability of annihilation
of a particle is $p$. Below we write down the possible
transitions rates related to a site and its two
nearest neighbors 
\[
\begin{array}{lll}
101 \to 111 & 1 &  \\
100 \to 110 & 1/2 &  \\
001 \to 011 & 1/2 &  \\
000 \to 010 & 0 &  
\end{array}
\hskip1cm
\begin{array}{lll}
111 \to 101 & p &  \\
110 \to 100 & p &  \\
011 \to 001 & p &  \\
010 \to 000 & p &  
\end{array}
\]

Using the equations (\ref{9}) and (\ref{10}), we write down the
entropy flux associated to the above transitions, that are related
to three given consecutive sites of the lattice.
For the transitions $101\to 111$ and $111\to 101$, the
fluxes are the same and given by
\beq
\psi_1 = \frac12(P_{101} - p P_{111})\ln\frac{1}{p},
\label{16a}
\eeq
where $P_{101}$ and $P_{111}$ are the probabilities of three
consecutive sites being in states 101 and 111, respectively.
The contribution coming from the transitions
$100\to 110$ and $110\to 100$ are equal and given by 
\beq
\psi_2 = \frac12(\frac12 P_{100} - p P_{110})\ln\frac{1/2}{p} 
\label{16b}
\eeq
whereas that associated to the transitions $001\to 011$ and 
$011\to 001$ are equal and given by
\beq
\psi_3 = \frac12(\frac12 P_{001} - p P_{011})\ln\frac{1/2}{p}.
\label{16c}
\eeq
The contribution coming from $010\to 000$ is
\beq
\psi_4 = - p (P_{010} - P_{000}).
\label{16d}
\eeq
We point out that this last expression is valid when $P_{010}$ is
nonzero. If $P_{010}$ vanishes then $\psi_4$ should be set equal to zero
even if $P_{000}$ is nonzero.
The total entropy flux associated to a triple of consecutive
sites is $\psi=2\psi_1+2\psi_2+2\psi_3+\psi_4$.

We have performed numerical
simulations on a chain with periodic boundary conditions and a
number of sites $L$ ranging from 100 to 2000. The simulation
was carried out as follows \cite{tome2015L}. At each time step,
a site is chosen at random and it is updated according to the rules
above. Our simulations is performed in such a way that we
do not allow the last particle to be annihilated \cite{tome2015L}.
Thus strictly speaking there is no absorbing states. However,
in the thermodynamic limit the system may reach a state
with vanishing density, which characterizes the absorbing
state. This indeed happens if the parameter $p$ is 
above its critical value $p_c$. That is, in the stationary state,
and in the thermodynamic limit,
the system is found in the state such that for $p\geq p_c$, the
density of particle $\rho$ vanishes whereas for $p<p_c$,
the system is in an active state with $\rho\neq0$. 

\begin{figure}
\centering
\epsfig{file=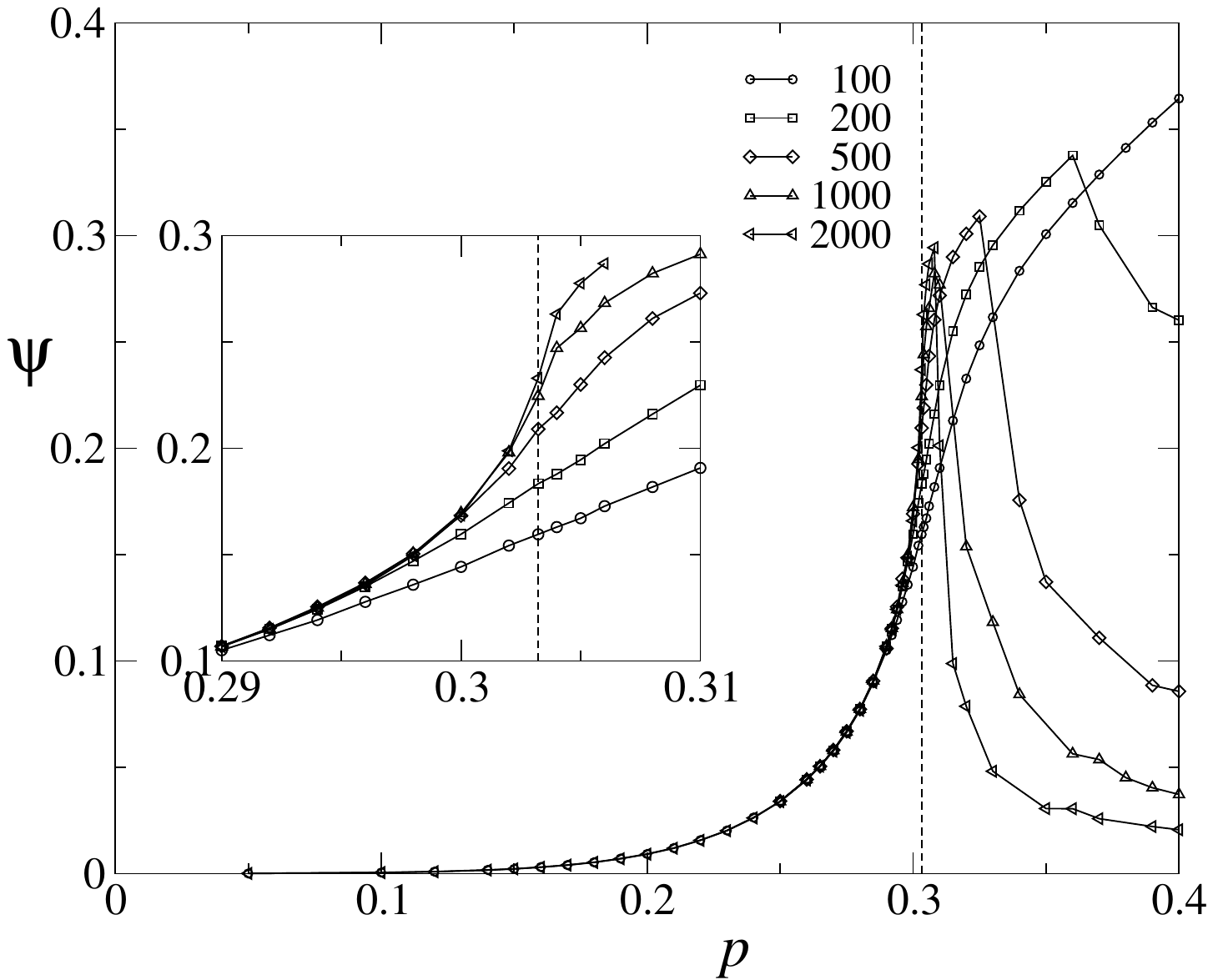,width=9cm}
\caption{Entropy production rate $\psi$ as a function of 
the annihilation parameter $p$, obtained from simulations
on chains of several sizes. The vertical dashed line indicates
the critical value $p_c=0.303228$. The inset shows an enlargement
around the critical point.}
\label{psi}
\end{figure}

The eight three-site probabilities that appear in the epressions
(\ref{16a}), (\ref{16b}), (\ref{16c}), and (\ref{16d})
are determined from the numerical simulations from
which we get $\psi_1$, $\psi_2$, $\psi_3$, 
$\psi_4$, and $\psi$. We recall that $\psi$ is also the rate
of entropy production, because we are considering the stationary
state. In figure \ref{psi} we show $\psi$ for several values of
$L$. As one increases the parameter $p$, the flux of entropy per
site $\psi$ also increases, reaches an inflection point, and
then decreases only after reaching a maximum. The inflection point
occurs near the critical point that takes place at $p_c=0.303228$
\cite{marro1999}.
From the inset of figure \ref{psi} we see that the slope
at the inflection point increases with the size of the lattice.

\begin{figure}
\centering
\epsfig{file=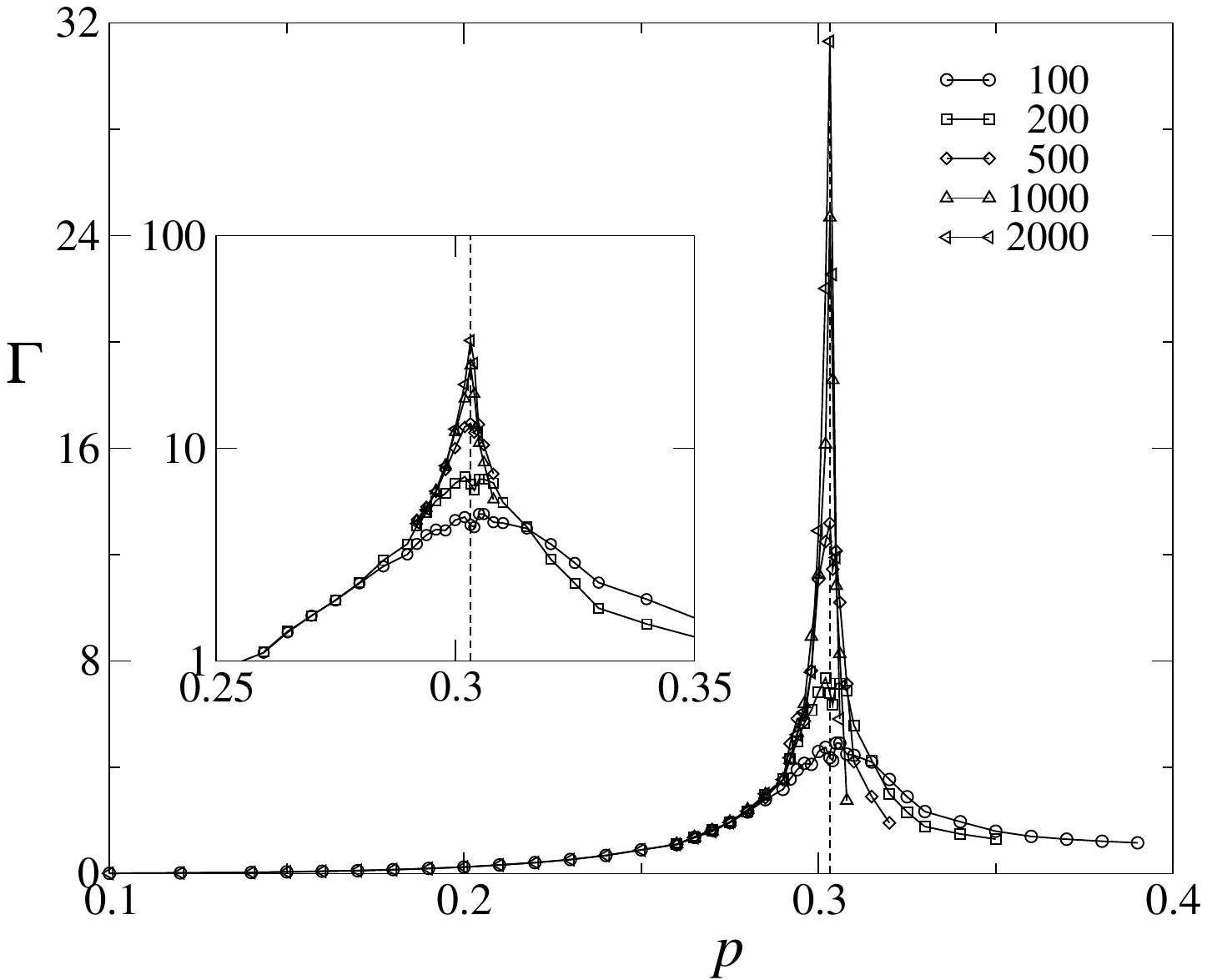,width=9cm}
\caption{Derivative of the entropy production rate $\Gamma=d\psi/dp$
as a function of the annihilation parameter $p$, obtained from the
data of figure \ref{psi}. The vertical dashed line indicates the
critical value $p_c=0.303228$. The inset shows a logarithm plot of
$\Gamma$ around the critical point.}
\label{deriv}
\end{figure}

These results indicates that in the thermodynamic limit, $\psi$
is finite at the critical point with a diverging slope at this point.
We thus assume that the singularity is described by 
\beq
\psi_0 - \psi \sim \varepsilon^{b},
\eeq 
where $\varepsilon=p_c-p$ is the deviation of the parameter $p$
from its critical value, and $0<b<1$. We remark that for $p>p_c$,
the entropy flux per site $\psi$ vanishes as $L\to\infty$
as can be inferred from the figure \ref{psi}, and because in this
limit the density vanishes. The singularity of the derivative
$\Gamma=d\psi/dp$ is described by
\beq
\Gamma \sim \varepsilon^{-a},
\eeq
where $a=1-b$, and diverges at the critical point,
as can seen in figure \ref{deriv}.

Using a finite size scaling we write
\beq
\Gamma_L = L^{a/\nu} f(L\varepsilon^\nu),
\eeq
where $f(x)$ is a scaling function, and $\nu$ is the critical
exponent associated to the correlation length.
From this scaling it follows that the slope at the
critical point increases with $L$ with the following
behavior
\beq
\ln \Gamma_L = \frac{a}{\nu} \ln L.
\eeq
From the numerical values of the slopes of $\psi$,
we obtain the following
value $a/\nu=0.67\pm0.05$. Using the value $\nu=1.097$ \cite{marro1999}
for the one-dimensional contact model we find $a=0.73 \pm0.05$, and 
$b=0.27\pm0.02$. 

The results above indicates that both the production of entropy 
$\psi$ and its derivative $\Gamma=d\psi/dp$ have a singular behavior
at the critical point that is described by a power law with exponents
$b$ and $a$. We point out that the numerical value of the exponent 
$b$ for the production of entropy is numerically equal to the
exponent $\beta$ associated to the order parameter $\rho$ of
the one-dimensional contact model which is $\beta=0.2765$
\cite{marro1999}. This result comes from the fact that the
entropy production is dominated by $P(010)$ which has a critical
behavior similar to the density of particles which is the
order parameter.

\section{Unidirectional transition rates}

Let us consider that the states 1 and 2 are connected
by a unidirectional transition rate $W_{21}>0$ from 1 to 2.
The contribution of this transition rate to $dP_1/dt$
is $-W_{21}P_1$ and to $dP_2/dt$ is $W_{21}P_1$, and the contribubtion
to $dS/dt$ is
\beq
W_{21}P_1\ln \frac{P_1}{P_2}. 
\label{38}
\eeq
Let us consider the case of a system such that at the
stationary state the difference $P_1-P_2$ is small.
In this case the expression above becomes 
\[
- W_{21}(P_2 - P_1).
\]
At the stationary state $P_1$ and $P_2$ does not depend on time
and $P_2\geq P_1$, and this contribution to $dS/dt$ is negative.
Since at the stationary state $dS/dt$ vanishes then there should
be positive contributions coming from the bidirectional 
transitions. If the transitions are all unidirectional then
at the stationary state $P_1=P_2$ and all probabilities are
equal. Taking into account that $dS/dt$ is equal to the entropy
production $\Pi$ {\it minus} the flux of entropy $\Psi$ then the term above
should be a part of the flux of entropy, that is, 
the contribution to $\Psi$ coming from the unidirectional
transition is 
\beq
\Psi_{21} = W_{21}(P_2-P_1). 
\label{39} 
\eeq
Summing this term with (\ref{38}), we find the contribution
of the unidirectional transition to the rate of entropy
production
\beq
\Pi_{21} = W_{21}P_1\ln \frac{P_1}{P_2} + W_{21}(P_2-P_1).
\eeq
which turns out to be positive definite, and
is our proposed expression (\ref{7}).

The above reasoning does not show that the form (\ref{39})
is necessary but shows that it is sufficient. In fact
any other form $F(P_2,P_1)$ such that $F\geq 0$ for
$P_2\geq P_1$ will serve as long as
\beq
P_1\ln \frac{P_1}{P_2} + F(P_2,P_1)\geq 0,
\eeq
for {\it any} $P_1$ and $P_2$. 
Writing $F(P_2,P_1) = P_1f(P_2/P_1)$
this condition becomes
\beq
f(x) \geq \ln x,
\eeq
supplemented by $f(1)=0$. Our choice is $f(x)=x-1$.
The choice made by Busiello, Gupta, and Maritan
\cite{busiello2020} is
\beq
\Psi_{21} = W_{21} P_1 \ln \frac{P_2}{P_1}.
\eeq
and corresponds to $f(x)=\ln x$. But this expression for
$\Psi_{21}$ makes the corresponding entropy production
$\Pi_{21}=0$ to vanish identically.

An essential property required for the production of entropy is
that it vanishes at thermodynamic equilibrium. As the
production of entropy is composed of a sum of terms
that are each one nonnegative then each term should
vanish. In the case of the Schnakenberg formula (\ref{6}),
the vanishing of each term implies that the transition rates
should fulfill the condition 
\beq
W_{ij}P_j^0 = W_{ji}P_i^0,
\eeq
where $P_i^0$ is the equilibrium distribution.
In the case of formula (\ref{7}), the condition is
\beq
P_j^0=P_i^0.
\eeq
Therefore if both types of transitions are present,
unidirectional and bidirectional, then the thermodynamic
equilibrium is possible if $W_{ij}=W_{ji}$. 

We show next that the production of entropy
is a minimum at the equilibrium. To this end we 
write $P_i=q_i +\delta p_i$ and replace in 
\beq
\Pi_{ij} = W_{ij}[q_j\ln\frac{q_j}{q_i} - (q_j-q_i)].
\eeq
Up to second order in $\delta p_i$, we find\beq
\delta \Pi_{ij} = W_{ij}[\delta p_j \ln\frac{q_j}{q_i} - \delta p_i (\frac{q_j}{q_i} -1)]
+ \frac12 W_{ij}q_j(\frac{\delta p_j}{q_j}-\frac{\delta p_i}{q_i})^2.
\eeq
If $q_i=P_i^0$ the terms linear in $\delta p_i$ vanish and we are left
with 
\beq
\delta \Pi_{ij} = \frac{W_{ij}}{2P_i^0}(\delta p_j- \delta p_i)^2,
\eeq
which shows that the entropy production is a minimum at equilibirum.
The value of $\Pi_{ij}$ at equilibrium also vanish as expected.

We remark that the additivity of the entropy flux follows
immediately from the expression (\ref{22}) and is implicit
is our numerical calculation. Notice that the plots in
figure \ref{psi} shows the entropy flux per site of the
lattice.

\section{Conclusion}

We have proposed an expression for the rate of entropy
production for the case in which the reverse transition rate 
vanishes but the forward transition rate is nonzero.
This expression is positive definite and the corresponding
entropy flux is linear in the probability distribution
allowing its calculation as an average over this distribution.
This expression allows us to determine the rate of entropy
production for stochastic dynamic systems whose some of
the transitions do not have the reverse. This is the 
case of models with with absorbing states such as the
contact process.

We have applied our expression to the
one-dimensional contact model and determined is
rate of entropy production at the stationary state.
We found that the rate of entropy per site is finite 
having a singularity at the critical point. 
The derivative of the production of entropy
with respect to the annihilation parameter is
also singular at the critical point, diverging at 
this point. The critical exponent associated to 
critical behavior of the rate of entropy production
is found to be numerically equal to the exponent $\beta$
associated to the critical behavior of the
particle density of the one-dimensional contact model.



\begin{thebibliography}{99}

\bibitem{kampen1981} N. G. van Kampen,
{\it Stochastic Processes in Physics and
Chemistry} (North-Holland, Amsterdam, 1981).

\bibitem{tome2015L} T. Tomé and M. J. de Oliveira,
{\it Stochastic Dynamics and Irreversibility}
(Springer, Cham, 2015).

\bibitem{schnakenberg1976} J. Schnakenberg,
Rev. Mod. Phys. 48, 571 (1976).

\bibitem{luo1984} J.-L. Luo, C. Van den Broeck, and G. Nicolis,
Z. Phys. B {\bf 56}, 165 (1984).

\bibitem{mou1986} C. Y. Mou, J.-L. Luo, and G. Nicolis,
J. Chem. Phys. {\bf 84}, 7011 (1986).

\bibitem{lebowitz1999} J. L. Lebowitz and H. Spohn,
J. Stat. Phys. {\bf 95}, 333 (1999).

\bibitem{maes2003} C. Maes and K. Neto\v{c}ný,
J. Stat. Phys. {\bf 110}, 269 (2003).

\bibitem{seifert2005} U. Seifert,
Phys. Rev. Lett. {\bf 95}, 040602 (2005).

\bibitem{tome2006} T. Tomé,
Braz. J. Phys. {\bf 36}, 1285 (2006).

\bibitem{zia2007} R. K. P. Zia and B. Schmittmann,
J. Stat. Mech. (2007) P07012.

\bibitem{andrieux2007} D. Andrieux and P. Gaspar,
J. Stat. Phys. {\bf 127}, 107 (2007).

\bibitem{harris2007} R. J. Harris and G. M. Schütz,
J. Stat. Mech. (2007) P07020.

\bibitem{gaveau2009} B. Gaveau, M. Moreau, and L. S. Schulman,
Phys. Rev. E {\bf 79}, 010102 (2009).

\bibitem{esposito2009} M. Esposito, K. Lindenberg,
and C. Van den Broeck, Phys. Rev. Lett. {\bf 102}, 130602 (2009).

\bibitem{szabo2010} G. Szabó, T. Tomé, and I. Borsos,
Phys. Rev. E {\bf 82}, 011105 (2010).

\bibitem{tome2010} T. Tomé and M. J. de Oliveira,
Phys. Rev. E {\bf 82}, 021120 (2010).

\bibitem{hinrichsen2011} H. Hinrichsen, C. Gogolin, and P. Janotta,
Journal of Physics: Conference Series {\bf 297}, 012011 (2011).

\bibitem{esposito2012} M. Esposito,
Phys. Rev. E {\bf 85}, 041125 (2012).

\bibitem{tome2015} T. Tomé and M.J. de Oliveira,
Physical Review E {\bf 91}, 042140 (2015).

\bibitem{tome2018} T. Tomé and M. J. de Oliveira
Journal of Chemical Physics {\bf 148}, 224104 (2018).

\bibitem{crochik2005} L. Crochik and T. Tomé,
Phys. Rev. E {\bf 72}, 057103 (2005).

\bibitem{tome2012} T. Tomé and M. J. de Oliveira,
Phys. Rev. Lett. {\bf 108}, 020601 (2012).

\bibitem{noa2019}
C. E. Noa, P. E. Harunari, M. J. de Oliveira, C. E. Fiore
Physical Review E 100, 012104 (2019).

\bibitem{tome2023} T. Tomé, C.E. Fiore, and M. J. de Oliveira,
Physical Review E {\bf 107}, 064135 (2023).

\bibitem{hawthorne2023}
F. Hawthorne, P. E. Harunari, M. J. de Oliveira, and C. E. Fiore
Entropy {\bf 25}, 1230 (2023).

\bibitem{marro1999} J. Marro and R. Dickman,
{\it Nonequilibrium Phase Transitions in Lattice Models}
(Cambridge University Press, Cambridge, 1999).

\bibitem{hinrichsen2000} H. Hinrichsen,
Advances in physics {\bf 49}, 815 (2000).

\bibitem{henkel2008} M. Henkel, H. Hinrichsen, and S. Lübeck,
{\it Non-Equilibrium Phase Transitions}
(Springer, Dordrecht,2008); volume 1.

\bibitem{murashita2014} Y. Murashita. K. Funo, and M. Ueda,
Phys. Rev. E {\bf 90} 042110 (2014).

\bibitem{ohkubo2009} J. Ohkubo,
J. Phys. Soc. Japn. {\bf 78}, 123001 (2009)

\bibitem{benavraham2011} D. ben-Avraham, S. Dorosz, and M. Pleimling,
Phys. Rev. E {\bf 84}, 011115 (2011).

\bibitem{zeraati2012} S. Zeraati, F.H. Jafarpour, H. Hinrichsen,
J. Stat. Mech.  (2012) L12001.

\bibitem{rahav2014} S. Rahav and U. Harbola,
J. Stat. Mech. (2014) P10044.

\bibitem{saha2016} B. Saha and S. Mukherji,
J. Stat. Mech. (2016) 013202.

\bibitem{pal2017} A. Pal and S. Rahav,
Phys. Rev. E, {\bf 96}, 062135 (2017).

\bibitem{busiello2020} D. M. Busiello, D. Gupta, and A. Maritan,
Phys. Rev. Res. {\bf 2}, 023011 (2020).

\bibitem{pal2021} A. Pal, S. Reuveni, and S. Rahav,
Phys. Rev. Research {\bf 3}, 013273 (2021). 

\bibitem{pal2021a} A. Pal, S. Reuveni, and S. Rahav,
Phys. Rev. Research {\bf 3}, L032034 (2021).

\end{thebibliography}
\end{document}